\documentclass[aps,pra,reprint,nofootinbib,twocolumn,superscriptaddress,showpacs,showkeys,longbibliography]{revtex4-1}
\usepackage{eurosym}
\usepackage{amsmath,amssymb,amstext}
\usepackage[usenames,dvipsnames]{color}
\usepackage{graphicx}
\usepackage{braket}
\usepackage{natbib}
\usepackage{comment}
\usepackage{dcolumn}
\usepackage[english]{babel}
\usepackage{wasysym}
\usepackage[colorlinks,bookmarks=false,citecolor=blue,linkcolor=red,urlcolor=blue]{hyperref}
\newcommand{\mathsym}[1]{{}}
\newcommand{\unicode}[1]{{}}

\begin{document}
\title{Nonequilibrium Landau-Zener Tunneling in  Exciton-Polariton Condensates}
\author{Xingran Xu}
\affiliation{Department of Physics, Zhejiang Normal University, Jinhua, 321004, China}
\affiliation{Shenyang National Laboratory for Materials Science, Institute of Metal Research, Chinese Academy of Sciences, Shenyang, 110016, China}
\affiliation{ School of Materials Science and Engineering, University of Science and Technology of China, Hefei, 230026, China}
\author{Zhidong Zhang}
\affiliation{Shenyang National Laboratory for Materials Science, Institute of Metal Research, Chinese Academy of Sciences, Shenyang, 110016, China}
\affiliation{ School of Materials Science and Engineering, University of Science and Technology of China, Hefei, 230026, China}
\author{Zhaoxin Liang}
\email{The corresponding author: zhxliang@gmail.com}
\affiliation{Department of Physics, Zhejiang Normal University, Jinhua, 321004, China}
\date{\today}
\begin{abstract}
For describing a coherent quantum two-level system driven by a linearly time-dependent detuning, the Landau-Zener model is routine to serve as a textbook model of its dynamics. Along this research line,  a particularly intriguing question is whether such framework can be extended to capture an intrinsic nonequilibrium nature for a quantum system with coherent and dissipative dynamics occurring on an equal footing. In this work, we are motivated to investigate the Landau-Zener problem of polariton condensates in a periodic potential under nonresonant pumping by using the driven-dissipative Gross-Pitaevskii equations coupled to the rate equation. Within the two-mode approximation, a nonequilibrium Landau-Zener model, characterized by  coherent and dissipative dynamics occurring on an equal footing, are derived.  Fundamentally different from the previous Landau-Zener model,  the total density of nonequilibrium Landau-Zener model general is not the conserved quantity anymore due to the dissipative nature. In surprise, the parameter regimes of the total density still being conserved can still be found. The motion of Hamiltonian of non-equilibrium Landau-Zener problem in phase space is further discussed, which is directly corresponding to the tunneling rate. The instability of the band structure  can also be studied by the curvatures in phase space and there may be two loops in the middle of the Brillouin zone. Detailed analysis on the non-equlibrium nature on the tunneling rate will open a new perspective toward understanding the Landau-Zener problem.
\end{abstract}
\maketitle
\section{Introduction}\label{section:one}

Adiabatic transitions at avoided level crossings play an essential role in many dynamical processes throughout physics and chemistry with potential applications of the quantum state preparation. The classical theoretical model studying adiabatic transitions is referred as to Landau-Zener (LZ) problem. In more detail, the dynamics of a quantum system are restricted to two quantum states coupled with a constant tunneling matrix element. A control parameter is swept
through the avoided level crossing at a constant velocity. The focus is on the final occupation probability of the two states. This  pure Landau-Zener problem was solved by Landau and Zener~\cite{Landau1932,Zener1932} independently. Then Wu and Niu~\cite{Wu2000,Wu2003} have extended the LZ model from the linear quantum system to nonlinear physical systems. Along this research line, a timely question is whether that framework of the LZ model can be extended to capture an intrinsic nonequilibrium nature for a quantum system with coherent and dissipative dynamics occurring on an equal footing.

Recently, exciton-polariton condensates attract many interests for their exceedingly light effective mass (the order of $10^{-5}$ the mass of free electrons) which can be used to realize Bose-Einstein condense(BEC) at room temperature~\cite{Rev0,Rev1,Rev2,Rev3}. It is a nonequilibrium system because the lifetime of polaritons is short and needs to be replenished by a pump laser. There are lots of  work discuss the steady states and elementary excitations of polaritons in both one and two components condensates theoretically and experimentally~\cite{excitation1,excitation2,excitation3,excitation4,ExcitationE1,ExcitationE2,Xu2017,Spinint2,Takemura2014,two1,two2}, meanwhile, nonlinear phenomena like oblique dark solitons, vortices, bright solitons and dark-bright solitons in dissipative system bring a new research field~\cite{nonlinear1,nonlinear2,nonlinear3,nonlinear4,nonlinear5,nonlinear6,nonlinear7,Pinsker2015,Pinsker2016}. Spontaneous oscillations in a microcavity polariton bosonic Josephson junction with strong imbalance of the population have been observed in a double-well ~\cite{Josephson2010,Abbarchi2013,Josephson2013,Josephson2014} which arise more interests in studying polaritons in periodic potential. The periodic potential in polariton has been realized by surface acoustic wave(SAW) and buried mesa array in experiments~\cite{Cerda2012,Krizhanovskii2013,Chestnov2016,gao2018}. The band structure of spinor polariton arises much attentions accounting for it is a natural dissipative two-dimensional material to study topological properties~\cite{Topo1,Topo2}, spin Hall effects~\cite{Hall1,Hall2}, the flat band in Lie lattice~\cite{Biondi2015,Leykam2018,Sun2018,Leykam2018a,Bingkun2018} etc. Besides, tuning the interaction of polaritons is demonstrated by using biexcitonic Feshbach resonance in recent experiments~\cite{Vladimirova2010,NP2014}. With these technologies, the adiabatic theory in polaritons can be investigated experimentally and theoretically. The influence of noise and modulational instability in polaritons have been discussed in Refs.~\cite{Bobrovska2015,Opala2018,Baboux:18}. However, they all consider the reservoir has no fluctuations and substitutes the steady-state of the reservoir to get two coupling equations.

In this work, we are motived to theoretically investigate the non-equilibrium LZ problems of a polariton BEC under nonresonant pumping trapped in a periodic potential. First, we introduce an effective Hamiltonian for polariton condensates in periodic potential by using the mean-field approach in one dimension. We just consider two modes of polaritons and the reservoir has fluctuation with the same period of the given potential and the adiabatic coefficient is added to the model to study the Landau-Zener theory in a dissipative system. As a result, we get a two-component Gross-Piteavskii(GP) equation coupled to a reservoir and the effective pumping of two components is the same in consideration.

Second, we present the steady states of the model both numerically and analytically. Particle number of the system may be not conserved along with time for the short lifetime of polaritons, however, we give three cases that the system can keep the number of polaritons by calculating $dn/dt$. The numerical results show the fluctuation of the reservoir can be set to a constant when two modes are unchanged with the time and we simplify our model to a two-level model.

Next, we study the Landau-Zener tunneling of the system. It is hard to define a tunneling probability of a dissipative system with a non-conserved particle number, so we just use the occupation of each states to describe this adiabatic process. We find fluctuation has a peak near $t=0$ which leads to the atom loss and presents how the interaction affects the occupation after tunneling.

Finally, we study the motion of the two-level dissipative LZ model in phase space obtained from the polariton system. The imaginary part of Hamiltonian is a periodic function along with the relative phase while the real part can have a new result for adjustable condensate density. When the pumping rate is far beyond the threshold, there may be two crossovers in the middle of the Brillouin zone as is famous for "swallowtails"~\cite{Chestnov2016}.

The emphasis and value of the present work are to provide a theoretical model, i.e., an extended LZ model in describing
the open quantum system with coherent and dissipative dynamics occurring on an equal footing capturing the key information of the nonequilibrium nature affecting adiabatic transitions at avoided level crossings. We remark that in the case of
vanishing the dissipation parameters, our model can be simplified into the coherent model which has been widely explored both theoretically
and experimentally in the context of the ultracold quantum gas~\cite{Wu2000,Wu2003} .
 We hope the model adopted in this work
can serve as a simple model to study adiabatic transitions at the avoided level crossing for a nonequilibrium quantum system.

The paper is organized as follows. In Sec.~\ref{section:two}, we introduce polariton in a periodic potential, which can be described by a dissipative GPE coupled to the rate equation of a reservoir under pumping. In Sec.~\ref{section:three}, we investigate the steady states of the model both numerically and analytically and simplify the problem into a two-level problem. In Sec.~\ref{section:four}, we evolute the model from the lower level at $t=-\infty$ to study the Landau-Zener tunneling. In Sec.~\ref{section:five}, we transform the Hamiltonian into phase space to find the motion of fixed points under different pumping rates. In Sec.~\ref{section:six}, we conclude with a summary of our main results and final remarks.

\section{Model}\label{section:two}

In this work, we are interested in an exciton-polariton BEC under non-resonant pumping in the presence of a periodic potential. At the mean-field level, the dynamics of the order parameter for the condensate labeled by $\psi$ can be well described by the driven-dissipative GP equation, i.e.,
\begin{eqnarray}
i\hbar\frac{\partial}{\partial t}\psi &=&-\frac{1}{2m}\left(\hbar\frac{\partial}{\partial x}-i\alpha t\right)^{2}\psi+V_{0}\cos\left(k_Lx\right)\psi\nonumber\\
&+&g\left|\psi\right|^{2}\psi+g_Rn_{R}\psi+i\frac{\hbar}{2}\left(Rn_{R}-\gamma_{C}\right)\psi,\label{psi}
\end{eqnarray}
with $m$ being the mass of the polariton, $g$ the interaction constant, $\gamma_C$ the decay rate of the polariton condensate, and $g_R$ characterizing the interaction between the condensate and reservoir.  $V_0$ in Eq. (\ref{psi}) is the strength of the periodic potential with the wavenumber of $k_L$ that can be controlled with a spatial quantization energy modulation of either the photonic or excitonic component~\cite{Winkler2016,AMO2016,Nalitov2017,gao2018}. The term of $\alpha=ma_c$ in Eq. (\ref{psi}) can be regarded as the vector potential gauge, which may be due to either the inertial force in the comoving frame of an accelerating lattice or the gravity force.

Equation (\ref{psi}) is coupled to an incoherent reservoir $n_R$, which is described by a rate equation, i.e.
\begin{equation}
\frac{\partial}{\partial t}n_{R}=P-\gamma_{R}n_{R}-R\left|\psi\right|^{2}n_{R},\label{eqnR}
\end{equation}
where $P$ is an off-resonant continuous-wave pumping rate, $\gamma_R$ is the dissipative rate of the reservoir and $R$ stands for the stimulated scattering rate of reservoir polaritons into the condensate.

Without both the periodic potential and the gauge potential, i.e. $V_0=0$ and $\alpha=0$ in Eq. (\ref{psi}), the steady-state under a continuous-wave and uniform
pumping can be obtained as $\psi_0=\sqrt{n_0}e^{-i(gn_0+g_Rn_R^0) t/\hbar}$ and  $n_R^0=\gamma_C/R$ with $n_0=\left(P-P_{\text {th}}\right)/\gamma_C$ and $P_{\text {th}}=\gamma_R\gamma_C/R$.

In the presence of periodic lattice potentials and the gauge potential, i.e. $V_0\neq 0$ and $\alpha\neq 0$ in Eq. (\ref{psi}),  there will exist the band structure in the polariton condensate,  where the longest lifetime characterizes the lowest-band top state. We are motivated to project Eqs. (\ref{psi}) and (\ref{eqnR}) into the
plane wave two-mode basis in the neighborhood of the Brillouin zone edge of $k=1/2$ and search for the solution in
the form
\begin{eqnarray}
\psi\left(x,t\right)&=&a\left(t\right)e^{ikx}+b\left(t\right)e^{i\left(k-1\right)x} \label{wf},\\
n_R\left(x,t\right)&=&n_{R}^{0}+2u\left(t\right)\cos x,\label{nRf}
\end{eqnarray}
with the reservior density having a periodic fluctuation described by the $u$ term in Eq. (\ref{nRf}). To simplify our calculation, we have set $\hbar=m=k_L=1$ and $V_0$ is relabbeled by  $v$. Then, by substituting Eqs.~(\ref{wf})-(\ref{nRf}) to Eqs. (\ref{psi}) and (\ref{eqnR}) and only collecting the coefficients of $e^{ikx}$ and $e^{i\left(k-1\right)x}$ for Eq. (\ref{psi}) and $e^{ix}$ and $e^{-ix}$ of Eq.~(\ref{eqnR}), we can obtain an extended LZ model  as follows
\begin{widetext}
\begin{eqnarray}
i\frac{\partial}{\partial t}\left(\begin{array}{c}
a\\
b\\
u
\end{array}\right)=\left(\begin{array}{ccc}
\emph{L}\left(k\right)+g\left(\left|a\right|^{2}+2\left|b\right|^{2}\right) & v/2 & \left(g_{R}+\frac{i}{2}R\right)b\\
v/2 & \emph{L}\left(k-1\right)+g\left(2\left|a\right|^{2}+\left|b\right|^{2}\right) & \left(g_{R}+\frac{i}{2}R\right)a\\
-\frac{iRn_{R}^{0}b^{*}}{2} & -\frac{iRn_{R}^{0}a^{*}}{2} & -i\left[\gamma_{R}+R\left(\left|b\right|^{2}+\left|a\right|^{2}\right)\right]
\end{array}\right)\left(\begin{array}{c}
a\\
b\\
u
\end{array}\right),\label{matrix}
\end{eqnarray}
\end{widetext}
with
\begin{equation}
\emph{L}\left(k\right)=\left[\frac{1}{2}(k-\alpha t)^{2}+g_{R}n_{R}^{0}+\frac{1}{2}i\left(Rn_{R}^{0}-\gamma_{C}\right)\right].
\end{equation}
Equation (\ref{matrix}) is the central result of this work, which describes a LZ model (i.e. Eq. (\ref{matrix})) for an intrinsically non-equilibrium quantum system
with coherent and dissipative dynamics occurring on an equal footing.  We also remark that, to our best knowledge, the non-equilibrium LZ problem described by Eq. (\ref{matrix})  is investigated for the
first time in the context of the polariton condensate.

To illustrate our model based on Eq. (\ref{matrix}) the analogies to and differences from the pure LZ model in Refs. \cite{Landau1932,Zener1932,Wu2000,Wu2003},
we first briefly demonstrate how Eq. (\ref{matrix}) can be simplified into the pure LZ model in the limiting case corresponding to $u=0$, $\gamma_C=0$, $R=0$, and
$g_R=0$. Here, we are interested in the adiabatic approach by taking $\alpha$ a small value. After only keeping the linear term of $\alpha$ in Eq. (\ref{matrix}), we can arrive at the nonlinear
LZ model~\cite{Wu2000,Wu2003} as follows
\begin{eqnarray}
i\frac{\partial}{\partial t}\left(\begin{array}{c}
a\\
b
\end{array}\right)=\left(\begin{array}{ccc}
-\alpha t+H_{\text{int}} & \frac{v}{2} \\
\frac{v}{2} &\alpha t-H_{\text{int}}
\end{array}\right)\left(\begin{array}{c}
a\\
b
\end{array}\right),\label{NLZ}
\end{eqnarray}
with $H_{\text{int}}=g(|b|^{2}-|a|^{2})$. For convenience, we have dropped out the average of the diagonal elements because it does not affect the evolution of the
probabilities. The total probability $|a|^2+|b|^2$ is conserved and is set to be $1$. If we further let $g=0$ in Eq. (\ref{NLZ}), the nonlinear LZ model will become to be the pure LZ model as it is expected~\cite{Landau1932,Zener1932}. We emphasize that in the both cases of pure LZ and nonlinear LZ, the total density labelled by $|a(t)|^2+|b(t)|^2=1$ is a conserved quantity, i.e. $d(|a(t)|^2+|b(t)|^2)/dt=0$. In high contrast, with considering the existence of an intrinsically non-equilibrium nature, the probability of $|a(t)|^2+|b(t)|^2$ immediately becomes to be a non-conserved quantity, representing the key physics of the non-equilibrium nature affecting the LZ problem.

\section{Steady states of Landau-Zener model with dissipation}\label{section:three}

As mentioned in Sec. \ref{section:two}, both the pure LZ and nonlinear LZ transitions are referred as to the coherent dynamical problems characterized with the total density
of $|a|^2+|b|^2$ being conserved. Fundamentally, the conserved total density is immediately violated [i.e. $d(|a|^2+|b|^2)/dt\neq 0$] for the introduction of the dissipation as shown in Eq. (\ref{matrix}).
Lacking the conserved total density brings the difficulty of defining the tunneling probability properly.  To defining the tunneling probability, our strategy is to take the value of $a$ (or $b$) in the limit of a long time with $a_\infty=\lim_{t\rightarrow \infty}a(t)$, which becomes to be time-independent, by solving the equations of motion of Eq. (\ref{matrix}). The goal of this section is to check whether there always exist the stationary states of our model in the considered parameter regimes. In this end, we are limited to the case of $\alpha=0$ in Eq. (\ref{matrix}) and investigate the long-time behavior of $a$ (or $b$) by numerically solving Eq. (\ref{matrix}) with different initial conditions.

\begin{figure}[h!]
\includegraphics[width=0.5\textwidth]{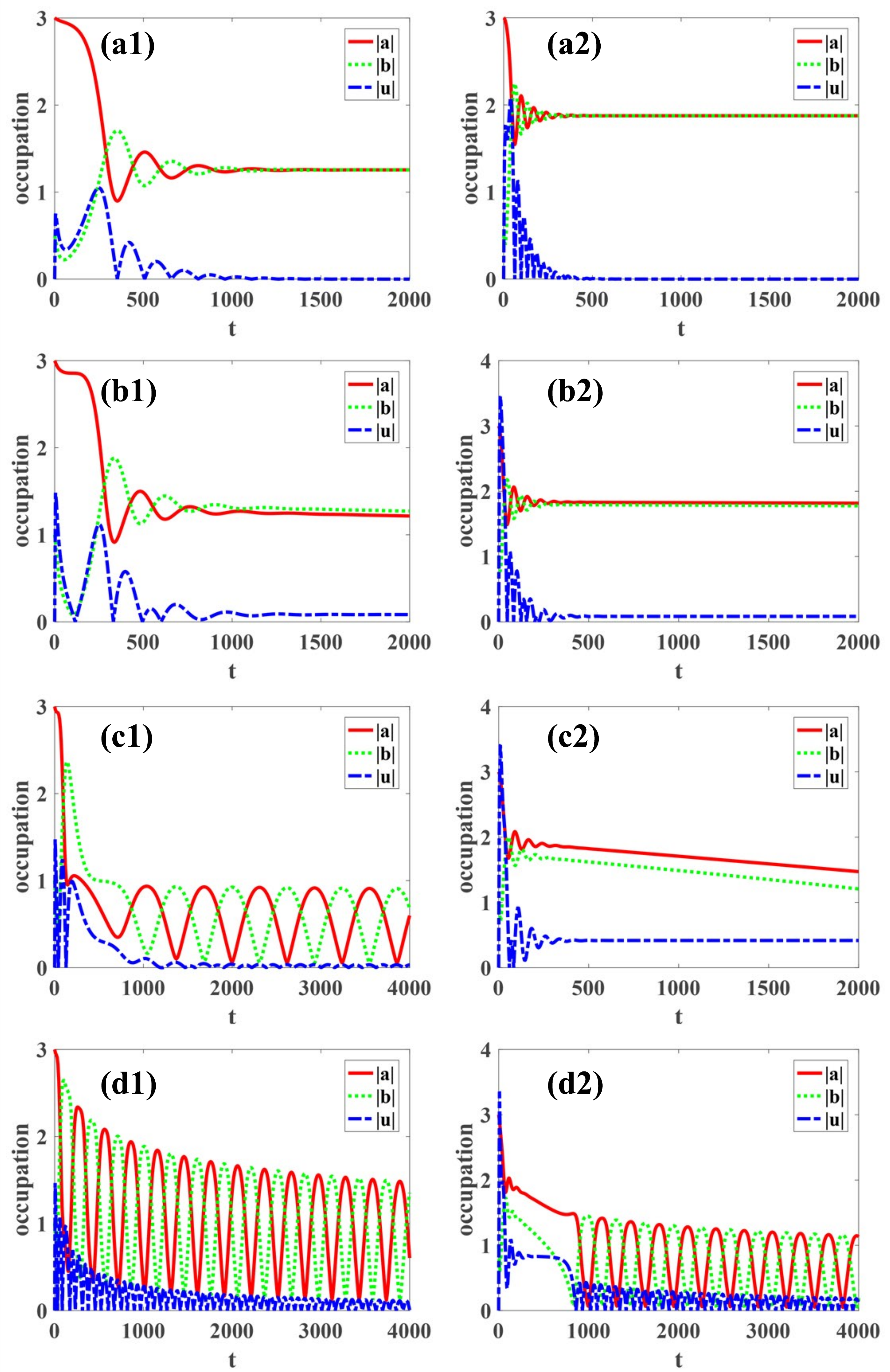}\\
\caption{Nonlinear dynamics of the Bloch mode amplitudes
governed by the model in Eq.~(\ref{matrix}). Parameters are used: $\alpha$=0, $\gamma_C$=0.33, $P$=35, $g$=0.006, $g_R$=2$g$, $R$=0.01, the first column $\gamma_R$=0.495, the second column $\gamma_R$=0.1, and (a1)-(a2): $v$=0; (b1)-(b2): $v$=0.002; (c1)-(c2): $v$=0.01; (d1)-(d2): $v$=0.02 . }\label{f1}
\end{figure}

We first choose the case of vanishing periodic strength $v=0$ and check whether our numerical results can recover the previous well-known results as a test of the validity of our numerical method.
As is illustrated in Figs.~\ref{f1} (a1) and (b1), for the case of the periodic potential strength $v=0$, the fluctuation of the reservoir can be set zero and the system can evolute into a stationary state with two components are the same no matter $\gamma_R<\gamma_C$ or $\gamma_R>\gamma_C$. If $u$ is zero, we can use the approach for reservoir density $n^0_R=P/(\gamma_R+R \left|\psi\right|^2)$. For vanishing $v$, the model is like a two-level system for ground and first excited state and this ansatz can also be used into two-dimension conditions for $b$, $a$ as a function of direction coordinate: $x$, $y$. Bloch mode dispersions of $p$, $\sigma$ and $\pi$ bands are observed in experiments \cite{gao2018}.

Then we proceed to consider how the periodic potential affects the stationary states by introducing $v\neq 0$. From Fig.~\ref{f1} (b1)-(b2) we can find the population of two components BECs can be different when $v$ raises and the fluctuation of the reservoir can be stable and nonzero. When $v/2$ is close to $g$, occupation of each double-well decreases and the fluctuation of reservoir changes with time and the occupation becomes unstable in Fig.~\ref{f1}(c1) with $\gamma_C<\gamma_R$, while in Fig.~\ref{f1} (c2) the occupation decreases even linearly with time and the fluctuation can still be stable. Finally, occupation of each well decreases with the change of fluctuation $\left|u\right|$, that procedure can keep a very long time as  shown in Figs.~\ref{f1}(d1)-(d2), while the fluctuation of reservoir becomes an oscillating function and vibrates in a high speed. The occupation in two wells will oscillate  and decrease when $v/2>g$, these phenomena have been observed in recent experiments~\cite{Josephson2010,Josephson2013}.

\begin{figure}[h!]
\includegraphics[width=0.5\textwidth]{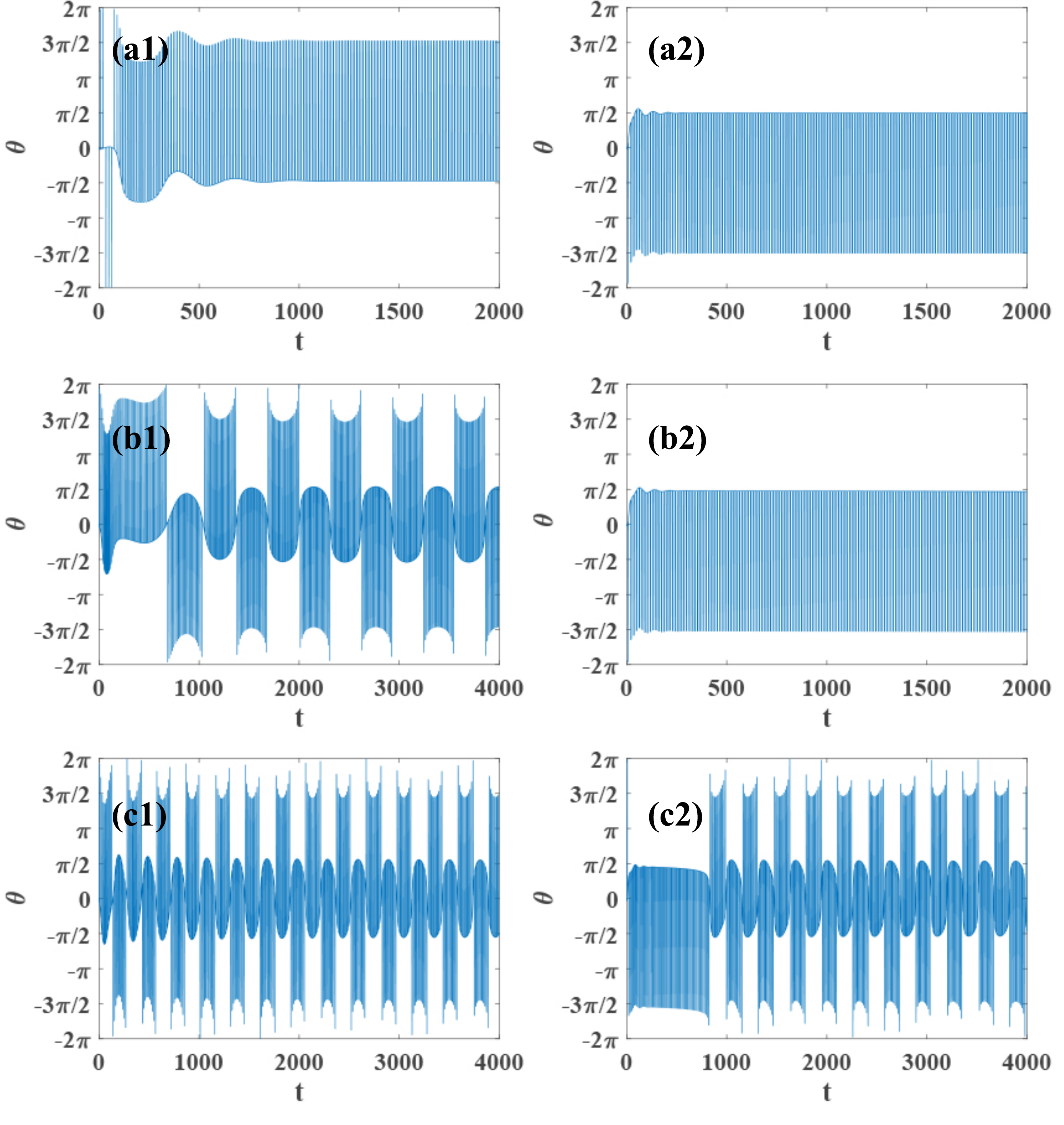}\\
\caption{Nonlinear dynamics of the Bloch mode relative phase $\theta=\theta_b-\theta_a$
governed by the model in Eq.~(\ref{matrix}). Parameters are used: $\alpha$=0, $\gamma_C$=0.33, $P$=35, $g$=0.006, $g_R$=2$g$, $R$=0.01, the first column $\gamma_R$=0.495, the second column $\gamma_R$=0.1, and (a1)-(a2): $v$=0.002; (b1)-(b2): $v$=0.01; (c1)-(c2): $v$=0.02 . }\label{f1phase}
\end{figure}

The long-time-independent stationary states in Fig.~\ref{f1} can be explained by calculating the time variation of the total density probability with the help of Eq. (\ref{matrix}) as,
\begin{equation}
\frac{d}{dt}\left(\left|a\right|^{2}+\left|b\right|^{2}\right)=Ru\left(a^{*}b+b^{*}a\right).\label{nq}
\end{equation}
From Eq. (\ref{nq}), the condition of the total density probability being conserved depends on two parameters: the fluctuation of reservoir $u$ and the relative phase between $a$ and $b$.
In such, there are three possible cases under which the total density probability still is a conserved quantity even with dissipation.
\begin{itemize}
\item \textbf{Case 1:} $u$=0, which indicates the fluctuation of reservoir is zero. The reservoir is a very large bath and the polariton cannot influence it.
\item \textbf{Case 2:} $|a|$ or $|b|$ is zero for one of the component vanishes. This means the particles are willing to stay in one well.
\item \textbf{Case 3:} The relative phase between $a$ and $b$ reads $\theta_a-\theta_b=\frac{\pi}{2}+j\pi$ with $j$ bing an integer.
\end{itemize}

In what follows, we plan to further check another possible case of the total density probability still being a conserved quantity even with dissipation.
First, we consider the system is still stable in the case of $u$ being a very small quantity compared to the reservoir density of $n^0_R$, although the total density probability
is not the conserved quantity any more according to Eq. (\ref{nq}) as shown in Fig.~\ref{f1}.  According to Eq.~(\ref{eqnR}), we can obtain the steady value of the reservoir as follows
\begin{equation}
n_R^0=\frac{P}{\gamma_R+R \left|\psi\right|^2}\approx\frac{P}{\gamma_R}\left(1-\frac{R}{\gamma_R}\left|\psi\right|^2\right),
\end{equation}
for $R\ll\gamma_R$. By substituting it to Eq.~(\ref{matrix}) as done by Ref.~\cite{Josephson2010},  we can obtain the analytical expression of $u$
\begin{equation}
u=-Rn_{R}^{0}\frac{ab^{*}+a^{*}b}{2\gamma_{R}+2R\left(\left|b\right|^{2}+\left|a\right|^{2}\right)},
\end{equation}
The density of  Eq. (\ref{nq}) becomes to ber
\begin{equation}
\frac{dn}{dt}=-R^2n_{R}^{0}\frac{\left(ab^{*}+a^{*}b\right)^2}{2\gamma_{R}+2R\left(\left|b\right|^{2}+\left|a\right|^{2}\right)}.
\end{equation}
In this work, we limit ourselves to be the parameter regimes of the relative phase between $a$ and $b$ reads $\theta_a-\theta_b=\frac{\pi}{2}+j\pi$ with $j$ bing an integer, in which
the total density is always a conserved quantity.
Within these parameter regimes, we substitute the form of $u$ to Eq.~\ref{matrix}, where the energy has a drop of $E_{\text{shift}}=1/8+g_{R}n_{R}^{0}$ with the steady reservoir $n_R^0=\gamma_C/R$, then we can re-write our model into the following form
\begin{widetext}
\begin{eqnarray}
i\frac{\partial}{\partial t}\left(\begin{array}{c}
a\\
b
\end{array}\right)=\left(\begin{array}{cc}
g\left(2\left|b\right|^{2}+\left|a\right|^{2}\right)-\left(\lambda+i\chi\right)\left|b\right|^{2}-\gamma/2 & v/2-\left(\lambda+i\chi\right)a^{*}b\\
v/2-\left(\lambda+i\chi\right)b^{*}a & g\left(2\left|a\right|^{2}+\left|b\right|^{2}\right)-\left(\lambda+i\chi\right)\left|a\right|^{2}+\gamma/2
\end{array}\right)\left(\begin{array}{c}
a\\
b
\end{array}\right)\label{matrix2}
\end{eqnarray}
\end{widetext}
with $\left(\lambda+i\chi\right)=\frac{\left(g_{R}+\frac{i}{2}R\right)Rn_{R}^{0}}{2\gamma_{R}+2Rn}$. Eq.~(\ref{matrix2}) is an effective model's Hamiltonian for our system and we can also define $\tilde{g}_{12}=2g-(\lambda+i\chi)$ for an effective interaction between the polaritons in different wells.

In Fig.~\ref{f1phase}, we investigate the relative phase of two Bloch modes is corresponding to our analytical result, the relative phase changes between $-\pi/2$ and $3\pi/2$ in Fig.~\ref{f1phase}(a1), while in strong pumping region change between $-3\pi/2$ and $\pi/2$. When the fluctuation oscillates with the time in Fig.~\ref{f1}, the relative phase is also oscillating around $\pi/2+j\pi$. There are some topological properties that keep the relative phase very important and can be used to lock phase. In Ref.~\cite{Josephson2010}, they observe a smoothed sawtooth phase evolution with a much smaller amplitude ($-0.3\pi$ to $0.3\pi$) which is very close to $-0.5\pi$ and $0.5\pi$ in our theory and the profile of population change with time is corresponding to ours. In Fig.~\ref{f1}(c2) and Fig.~\ref{f1phase} (b2) the population of polaritons cannot reach a balance in strong pumping region ($\gamma_R<\gamma_C$), because the imaginary part of elementary excitation can be larger than zero and the relative phase is not exactly between $-3\pi/2$ and $\pi/2$ but it is also a stable relative phase leading to linear decay.

\section{Landau-Zener Tunneling}\label{section:four}
In this section, we will pay attention to the adiabatic process by evolving a state from $t=-\infty$ to $\infty$ numerically. There are lots of work that has been done in adiabatic theory~\cite{Wu2000,Wu2003,Wang2018,Bobrovska2015}, but for the polariton system, there are some new results. In the past, the initial state is set to the lower state and the upper state can be calculated as tunneling probability for the conserved population, however, it is difficult to define the tunneling opportunity in an open system. We just focus on the occupation of each state before and after tunneling for the atom loss of the system. Considering the linear of $t$ for Hamiltonian is dependent on time and changes slowly in a nonlinear polariton system. In Sec.\ref{section:three}, we  investigate the occupation of each state will oscillate and decrease when $v$ is large enough. More details about tunneling will be discussed in this section.

\begin{figure}[h!]
\includegraphics[width=0.5\textwidth]{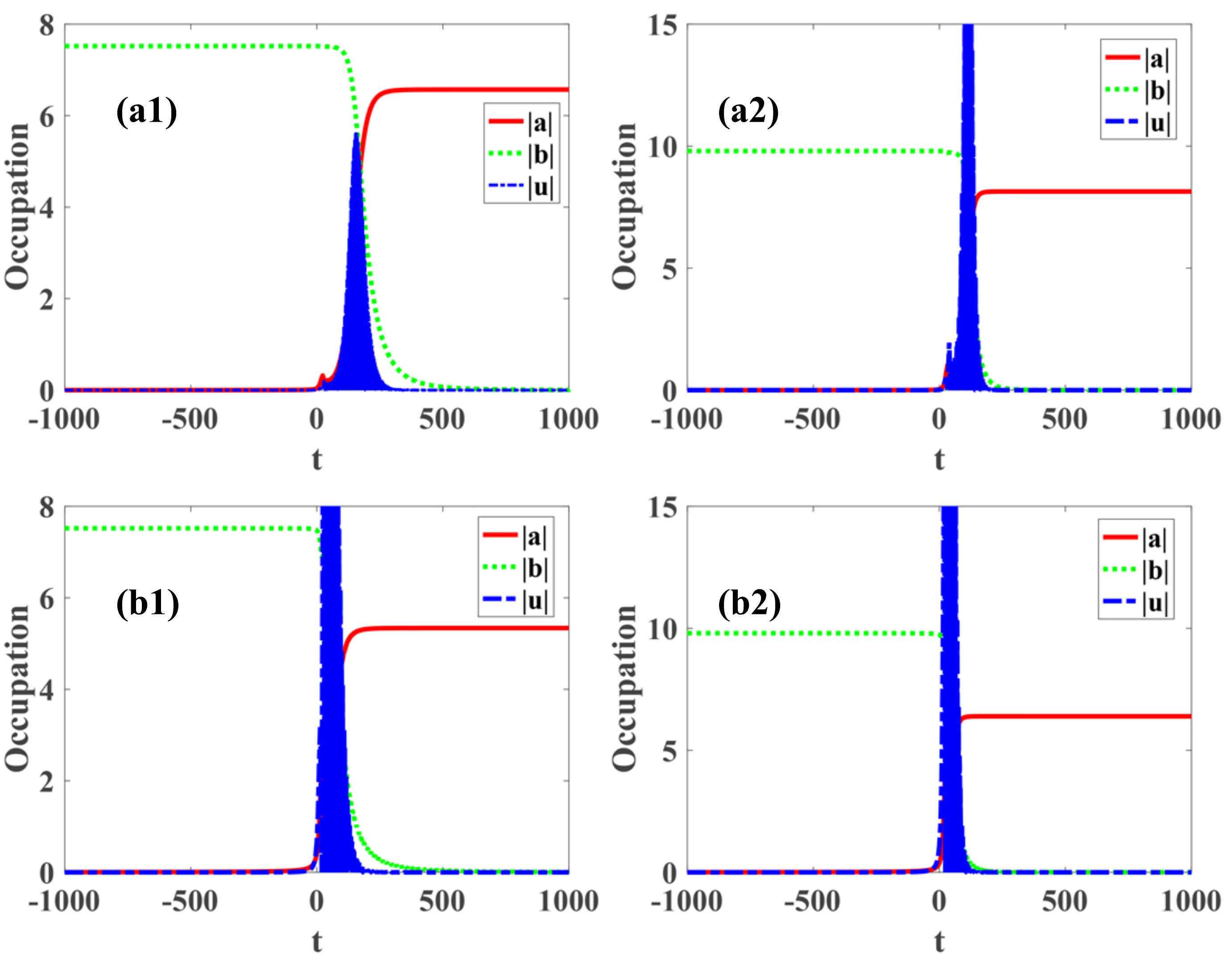}\\
\caption{Numerical results for the time evolution of Eq. (\ref{matrix}) with interaction strength $g$=0.006. Parameters are used: $\alpha$=0.02, $g_R$=2$g$, $P$=35, $\gamma_C$=0.33, $R$=0.01; and (a1): $v$=0.002, $\gamma_R$=0.495; (b1): $v$=0.002, $\gamma_R$=0.1; (a2):$v$=0.02, $\gamma_R$=0.495; and (b2): $v$=0.02, $\gamma_R$=0.1. \label{f2}}
\end{figure}
As is vividly shown in Fig.~\ref{f2}, we evolute the initial state from $\left(a,b,u\right)=\left(0,\sqrt{n_0},0\right)$ and find the fluctuation has a sudden change at $t=0$ from zero to nonzero. Two levels will have a crossover near $t=0$ and the initial occupation will vanish when $t>0$ for the fluctuation of the reservoir. Comparing Figs.~\ref{f2}(a1)-(b1) with (a2)-(b2), the peak of fluctuation in weak pumping region($\gamma_R>\gamma_C$) is lower than that in strong pumping region ($\gamma_R<\gamma_C$). We can also find larger depth of periodic potential can reduce more occupation in final states after tunneling as is shown $v=0.02$ in Fig.~\ref{f2}(b1) compared to $v=0.002$ in Fig.~\ref{f2}(a1) if we fix interaction strength. If it is a double-well system to study the Josephson junction between two wells, the oscillations have a short lifetime for particle number decreases after tunneling if the temperature is beyond zero($\alpha>0$).
\begin{figure}[h!]
\includegraphics[width=0.5\textwidth]{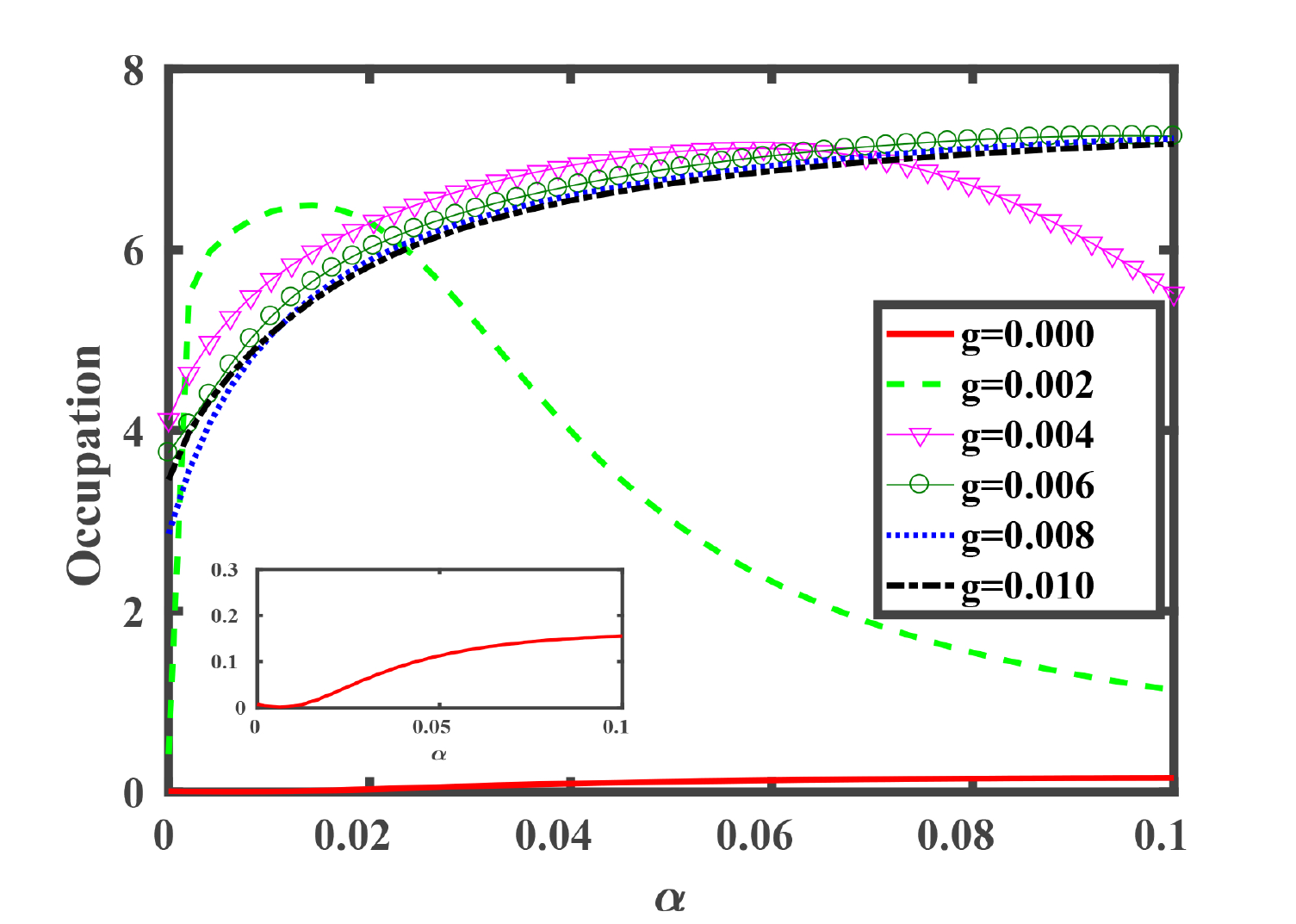}\\
\caption{Numerical results for the tunneling occupation as a function of $\alpha$ using Eq. (\ref{matrix}) with the same initial states $\left(a,b,u\right)$=$\left(0,\sqrt{n_0},0\right)$. Parameters are used: $g_R$=2$g$, $P$=35, $\gamma_C$=0.33, $\gamma_R$=0.495$, R$=0.01 and $v$=0.01. \label{f3}}
\end{figure}

It is known to all interaction strength $g$ can change the tunneling probability from zero to finite value for an adiabatic evolution and it will breakdown the Bloch oscillation. In recent experiments, the interaction of polariton can be adjusted by Feshbach resonance~\cite{NP2014}, so we can use the numerical calculation to find the rule of tunneling occupation along with adiabatic coefficients. The inset in Fig.~\ref{f3} is the enlarged drawing of the condition $g=0$ and we can see branch point is off the real axis leading to a transition occupation vanishing exponentially in the adiabatic limit~\cite{Wu2000,Wu2003}. The result is completely according to the close system for we set $g_R=2g$ and there is no interaction between the reservoir and the polaritons but the dissipative $\gamma_C$ leads to the particle loss. Here, we just consider $P\gg P_{th}$ in Fig.~\ref{f3}. When interaction $g$ is beyond zero, occupation after tunneling will decrease after reaching a peak for the fluctuation of the reservoir which is a result of competition between Landau-Zener driving and relaxation\cite{Nalbach2009,Javanbakht2015,Huang2018,Malla2018}. Besides, the increase of nonlinear interaction between the reservoir and polaritons makes the system stable again, which leads to dynamical stability even though the accelerate $\alpha$ is much larger. The competition between dissipative coefficients and the nonlinear interaction of reservoir and polaritons bring a powerful tool to achieve a new equilibrium.

\section{The motion of Hamiltonian in phase space}\label{section:five}

In this section, we focus on the motions of the polariton system in phase space. Here, we consider two components $a=ae^{i\theta_{a}}$, $b=be^{i\theta_{b}}$, population difference $s=\left|b\right|^{2}-\left|a\right|^{2}$, total population $n=\left|b\right|^{2}+\left|a\right|^{2}$, the relative phase $\theta=\theta_{b}-\theta_{a}$, and the product of two components $\left|ab\right|=\sqrt{n^{2}-s^{2}}/2$.
\begin{figure}[h!]
\includegraphics[width=0.45\textwidth]{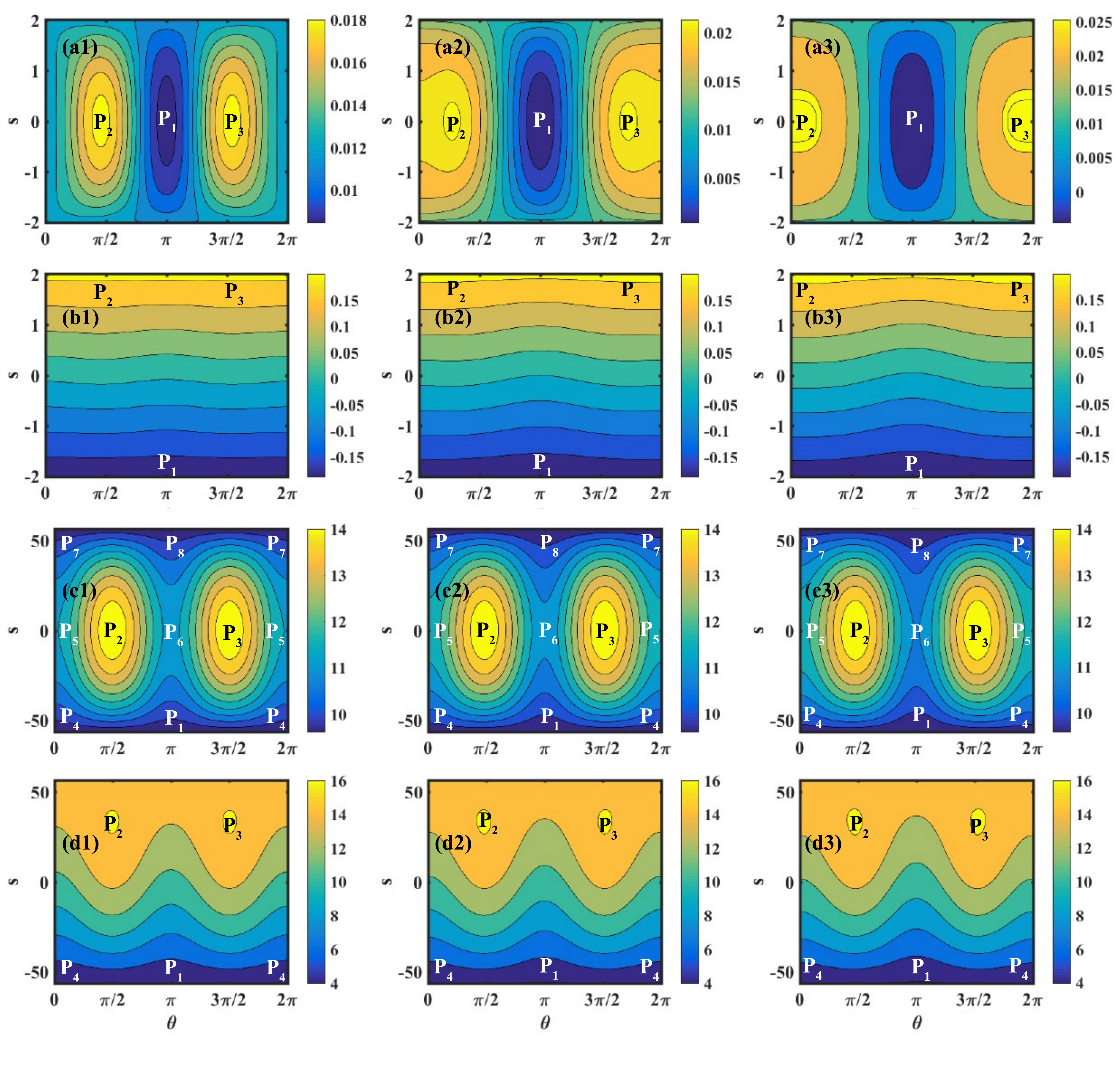}\\
\centering
\caption{Evolution of the phase space motions of the Hamiltonian system with interaction strength $g$= 0.006. Parameters are used: $\gamma_R$=0.496, $\gamma_C$=0.33, $R$=0.01, $g_R$=$2g$, and (a1)-(a3): $\gamma$=0, $P$=17; (b1)-(b3): $\gamma$=0.2, $P$=17; (c1)-(c3): $\gamma$=0, $P$=35; (d1)-(d3): $\gamma$=0.2, $P$=35; and the interaction for first column: $v$=0.002, the second column $v$=0.01 and the third column $v$=0.015.}\label{f4}
\end{figure}
From Eq.(\ref{matrix2}), the total
energy of the system can be cast into an effective Hamiltonian~\cite{Zobay2000,liu2003}:
\begin{eqnarray}
E&=&g\frac{3n^{2}-s^{2}}{4}+\gamma s/2+v\cos\theta\sqrt{n^{2}-s^{2}}/2\nonumber\\
&-&\left(\lambda+i\chi\right)\frac{n^{2}-s^{2}}{2}\cos^{2}\theta.
\end{eqnarray}
with $\gamma=\alpha t$. The energy is only corresponding to $n, s, \theta$ and the imaginary part of $E$ is always a periodic function to $\theta$, so we just consider the real part. The fixed
points of the effective Hamiltonian correspond to the eigenstates
of the nonlinear two-level system and the extreme points require :
\begin{eqnarray}
\frac{\partial}{\partial\theta}E &=&\left(n^{2}-s^{2}\right)\lambda\sin2\theta-\frac{\sqrt{n^{2}-s^{2}}}{2}v\sin\theta,\label{dtheta}\\
\frac{\partial}{\partial s}E &=&\frac{\gamma}{2}-\frac{gs}{2}+2\lambda s\cos^{2}\theta-\frac{sv\cos\theta}{2\sqrt{n^{2}-s^{2}}},\label{ds}\\
\frac{\partial}{\partial n}E &=&\frac{3gn}{2}+\frac{nv\cos\theta}{2\sqrt{n^{2}-s^{2}}}\nonumber\\
&-&\cos^{2}\theta\left[2\lambda n+\left(n^{2}-s^{2}\right)\frac{d\lambda}{dn}\right],\label{dn}
\end{eqnarray}
with $\frac{d}{dn}\lambda=-\frac{g_{R}Rn_{R}^{0}R}{2\left(\gamma_{R}+Rn\right)^{2}}$. From Eq.~(\ref{dtheta}), the fixed points only appear when $\sin\theta=0$ or $s=\pm n$, or $\cos\theta=\frac{v}{4\lambda\sqrt{n^{2}-s^{2}}}$. The solution is $s=\pm n$, or $\theta=0$, $\pi$ or $\theta=\arccos\left(\frac{v}{4\lambda\sqrt{n^{2}-s^{2}}}\right)$. Here we have a new solution if $\left|\frac{v}{4\lambda\sqrt{n^{2}-s^{2}}}\right| \leq1$ and
$n$ represents the density of polariton, so it must be a positive number and can be controlled by pump rate $P$.

First, we want to focus on the pumping rate is near the threshold $P_{th}$ and how the adiabatic process influences the phase space of the system. As is shown in Fig.~\ref{f4} (a1)-(a3), there are only three fixed points: one is at $\theta=\pi$ and others appear near $\theta=\pi/2$ and $\theta=3\pi/2$, along with the increase of $v$ , fixed points go to the border of phase at adiabatic process with $\gamma=0$. In Fig.~\ref{f4}(a3) fixed points $\text{P}_2$ and $\text{P}_3$ are the same. When $\gamma$ is beyond zero, fixed point $\text{P}_1$ goes to the border $s=-n$ and points $\text{P}_{2}$ and $\text{P}_3$ go to the border $s=n$ as is shown in Fig.~\ref{f4}(b1)-(b3).

Polariton system's particle number can be adjusted by pumping rate and  if we set $P\gg P_{th}$, the system will have different results. There are 8 fixed points in phase space: $\text{P}_1$, $\text{P}_6$, and $\text{P}_8$ appear at $\theta=\pi$, meanwhile, $\text{P}_2$ and $\text{P}_3$ appear at $\theta=\pi/2$ and $3\pi/2$ and other points appear at $\theta=0$ as is shown in Fig.~\ref{f4}(c1)-(c3). Point $\text{P}_5$ or point $\text{P}_6$ is the saddle point can annihilate itself by colliding with
$\text{P}_1$ as R changes slowly, leading to the breakdown of
adiabaticity of the tunneling~\cite{liu2003}. Furthermore, $\text{P}_5$ and $\text{P}_6$ will also annihilate with $\text{P}_4$ and $\text{P}_1$, so in this situation, there may be two loops at $k$=1/2 as are reported in Ref.~\cite{Chestnov2016}. There are two saddle points when $v/2<g$, because we need to compare $gn$ with $v/2$ and in our region $gn>>v/2$ for $P\gg P_{th}$.
\section{Conclusion}\label{section:six}

In summary, we obtain an effective Hamiltonian of polaritons in a periodic potential using two modes of approaches under nonresonant pumping. In the past, fluctuations are not under consideration for it is much smaller than the reservoir, but it plays a very important role in the tunneling process. The steady states of the system provide the relative phase can only be $\pi/2+j\pi$ in Josephson junction and atoms will lose after tunneling for fluctuations have a sudden peak near $t=0$. The numerical results of the tunneling process agree with  Ref.~\cite{Josephson2010} very well. If we set the fluctuation of the reservoir to a constant, the simplified model is similar to the model discussed in Ref.~\cite{gao2018}. Last but not least, the motion of Hamiltonian in phase space reveals two loops in the band structure which is obtained in Ref.~\cite{Chestnov2016} with different methods.
\section{Acknowledge}\label{section:seven}

We thank Y. Xue, Ying Hu, J. Liu, and B. Wu for stimulating discussions. This is supported by the National Natural Science Foundation of China (Grant No.11604300) and Key Projects of the Natural Science Foundation of China (Grant No. 11835011). Z. D. Z. is supported by the NSFC of China (Grant No. 51331006).
\bibliography{myr}
\end{document}